\documentclass[reprint,
superscriptaddress,
showpacs,
 amsmath,amssymb,
 aps,
pra,
latexsym,mathtools
]{revtex4-1}

\usepackage{graphicx}% Include figure files
\usepackage{dcolumn}% Align table columns on decimal point
\usepackage{bm}% bold math
\usepackage{color}
\usepackage[bottom]{footmisc}
\usepackage{ulem}
\usepackage{setspace}
\usepackage[nottoc]{tocbibind}
\usepackage{blindtext}
\usepackage{titlesec}
\usepackage{cancel}

\usepackage[mathlines]{lineno}% Enable numbering of text and display math
\begin{document}

%\preprint{APS/123-QED}

\title{Engineered two-photon dissipative confinement of a Kerr-cat qubit using SISIS quantum circuit refrigerator
}% Force line breaks with \\

\author{Shumpei Masuda}
\affiliation{
National Institute of Advanced Industrial Science and Technology (AIST),
Tsukuba 305-8565,Japan
}
\author{Tsuyoshi Yamamoto}
\affiliation{Univ. Grenoble Alpes, CEA, Grenoble INP, IRIG, PHELIQS, 38000 Grenoble, France}
\affiliation{Institute for Solid State Physics, the University of Tokyo, Kashiwa, Chiba 277-8581, Japan}
\author{Shuji Nakamura}
\affiliation{Global Research and Development Center for Business by Quantum-AI Technology (G-QuAT),
National Institute of Advanced Industrial Science and Technology (AIST),
Umezono 1-1-1, Tsukuba 305-8565, Japan}
\author{Daichi Sugiyama}
\affiliation{Department of Physics, Graduate School of Science,
Tokyo University of Science,
1-3 Kagurazaka, Shinjuku-ku, Tokyo 162-8601, Japan}
\affiliation{Research Institute for Science and Technology,
Tokyo University of Science,
1-3 Kagurazaka, Shinjuku-ku, Tokyo 162-8601, Japan}
\author{Akiyoshi Tomonaga}
\affiliation{Global Research and Development Center for Business by Quantum-AI Technology (G-QuAT),
National Institute of Advanced Industrial Science and Technology (AIST),
Umezono 1-1-1, Tsukuba 305-8565, Japan}
\date{\today}% It is always \today, today,

\begin{abstract}
Kerr-cat qubits realized in periodically driven superconducting nonlinear resonators are a promising platform for quantum information processing with biased noise.
Pure dephasing in such systems induces leakage out of the qubit subspace, motivating the use of quantum circuit refrigeration (QCR) to remove excess excitations.
While conventional superconductor--insulator--normal-metal--insulator--superconductor (SINIS)-based QCRs can suppress leakage via single-photon absorption, they also enhance QCR-induced phase-flip errors.
Here we investigate a QCR based on a superconductor--insulator--superconductor--insulator--superconductor (SISIS) junction coupled to a Kerr parametric oscillator (KPO).
We show that a SISIS-based QCR can operate in a regime where single-photon processes are suppressed while two-photon absorption dominates.
As a result, the proposed SISIS-based QCR achieves strong suppression of dephasing-induced leakage while substantially reducing the increase in phase-flip errors associated with QCR operation.
These results demonstrate that the proposed SISIS-based QCR provides an effective approach for mitigating leakage while limiting QCR-induced phase-flip errors in Kerr-cat qubits.
\end{abstract}

%\pacs{02.30.Yy, 37.90.+, 67.85.Fg, 03.75.Lm}% PACS, the Physics and Astronomy
\maketitle

%\tableofcontents

%\tableofcontents

%\section{THEORY}
\section{Introduction}
Among bosonic-qubit architectures, Kerr-cat qubits realized in superconducting nonlinear resonators have emerged as a promising platform for quantum information processing.
By encoding logical information in metastable coherent states, these qubits exhibit a strongly biased error channel, in which bit-flip errors are exponentially suppressed relative to phase-flip errors~\cite{Cochrane1999,Goto2016,Puri2017b,Tuckett2019,Ataides2021}.
This bias enables hardware-efficient quantum error-correction schemes tailored to asymmetric noise and has motivated extensive theoretical and experimental studies of cat-qubit-based quantum computation~\cite{Ataides2021,Darmawan2021}.
Recent experiments have demonstrated coherence times exceeding 1 ms and gate fidelities above 99\% in stabilized cat qubits~\cite{Grimm2020,Puri2020,He2023}.

A leading physical implementation of Kerr-cat qubits is the Kerr parametric oscillator (KPO), a parametrically driven superconducting resonator with Kerr nonlinearity~\cite{Meaney2014,Wang2019,Goto2019,Iyama2023,Yamaguchi2024,Hoshi2025}.
In a KPO, the interplay between parametric driving and Kerr nonlinearity stabilizes cat states formed from coherent states of the resonator.
Related implementations based on nonlinear circuit elements such as SNAILs have also demonstrated cat-state stabilization and high-fidelity operations~\cite{Grimm2020,Frattini2022,He2023}.
Beyond fault-tolerant quantum computation, Kerr-cat qubits provide a versatile platform for quantum annealing~\cite{Goto2016,Nigg2017,Puri2017,Zhao2018,Onodera2020,Goto2020a,Kewming2020,Kanao2021,Yamaji2022,Yamaji2025},
Boltzmann sampling~\cite{Goto2018},
and the exploration of nonequilibrium phenomena, including dissipative phase transitions~\cite{Dykman2018,Rota2019,Kewming2022} and quantum chaos~\cite{Milburn1991,Wielinga1993,Hovsepyan2016,Goto2021b}.

Despite the biased-noise advantage of Kerr-cat qubits, KPOs remain vulnerable to pure dephasing of the resonator.
Dephasing-induced frequency fluctuations can drive the system out of the computational subspace, resulting in leakage into higher excited states~\cite{Yamaji2022,Masuda2025}.
Such leakage errors are difficult to correct within conventional quantum-error-correction protocols because they occur outside the encoded qubit subspace.
To suppress leakage, several engineered-dissipation schemes have been proposed, including approaches based on two-photon loss~\cite{Touzard2018,Puri2020} and frequency-selective dissipation~\cite{Putterman2022}.
However, to our knowledge, experimental demonstrations of direct leakage suppression in KPO- or SNAIL-based Kerr-cat qubits using engineered dissipation remain limited.

Quantum circuit refrigeration (QCR), based on photon-assisted quasiparticle tunneling in microscopic junctions~\cite{Tien1963,Devoret1990,Girvin1990,Averin1990,Pekola2010}, provides a versatile approach to engineering dissipation in superconducting circuits.
QCRs based on normal-metal--insulator--superconductor (NIS) junctions have been experimentally demonstrated as controllable dissipation sources for superconducting resonators and qubits~\cite{Tan2017,Silveri2017,Silveri2019,Hsu2020,Hsu2021,Yoshioka2021}, and have recently been applied to fast qubit reset~\cite{Sevriuk2022,Yoshioka2023,Nakamura2025}.
In these devices, dissipation is predominantly mediated by single-photon absorption processes.

Recently, a QCR architecture based on superconductor--insulator--normal-metal--insulator--superconductor (SINIS) junctions was proposed to cool KPOs via single-photon absorption~\cite{Masuda2025}.
This approach can suppress dephasing-induced leakage by relaxing population from higher excited states back into the qubit subspace.
However, because single-photon processes change the parity of KPO states, they inevitably induce phase-flip errors within the computational subspace~\cite{Puri2017}.
Consequently, stronger cooling is generally accompanied by an increase in QCR-induced phase-flip errors.
Furthermore, the broad density of states of the normal-metal island favors single-photon tunneling, making multi-photon processes ineffective as a cooling resource.

In this work, we propose an alternative QCR architecture based on a superconductor--insulator--superconductor--insulator--superconductor (SISIS) junction capacitively coupled to a KPO.
Owing to the sharp superconducting density of states of the island, photon-assisted tunneling processes in SISIS junctions differ qualitatively from those in SINIS junctions.
In particular, we identify a bias-voltage regime in which single-photon processes are suppressed while two-photon absorption processes become dominant.
As a result, quasiparticle tunneling predominantly induces transitions within the same-parity manifold of the KPO, enabling efficient removal of dephasing-induced leakage while suppressing QCR-induced phase-flip errors.

%We theoretically analyze quasiparticle-tunneling-induced transitions in a driven KPO and evaluate the performance of the SISIS-based QCR in comparison with the SINIS-based approach.
%Our results show that exploiting two-photon tunneling processes enables strong suppression of leakage while largely preserving the biased error structure of Kerr-cat qubits.
%We focus on the regime where quasiparticle tunneling is the dominant transport mechanism, as detailed in Sec.~\ref{System}.

\section{Mechanism}
\label{Mechanism}
\begin{figure}[!]
\begin{center}
\includegraphics[width=7.5cm]{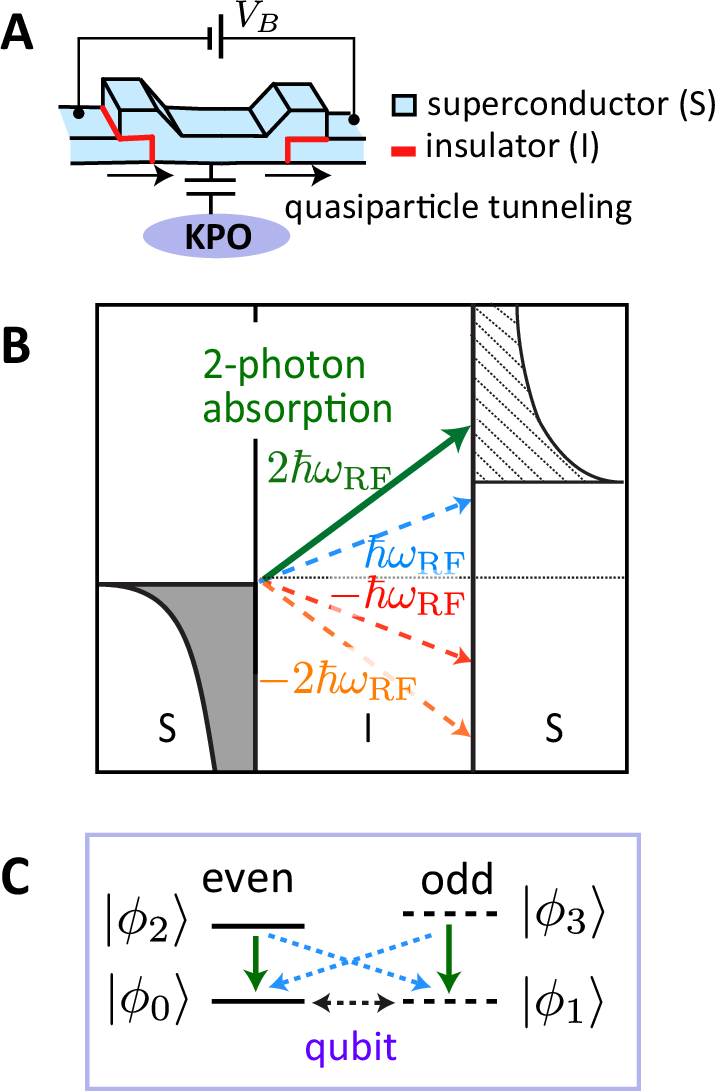}
\end{center}
\caption{
(A) Schematic of a SISIS-junction-based QCR capacitively coupled to a KPO.
A bias voltage $V_B$ is applied across the SISIS junction.
Horizontal arrows indicate quasiparticle tunneling processes.
(B) Energy diagrams for photon-assisted quasiparticle tunneling in a SIS junction in a bias regime where two-photon absorption remains allowed, whereas other photon-assisted tunneling processes are suppressed.
The black solid curves represent the superconducting density of states.
Colored and shaded regions indicate occupied and unoccupied states, respectively.
The arrows indicate two-photon absorption (green), single-photon absorption (blue), single-photon emission (red), and two-photon emission (orange).
Dashed arrows denote processes forbidden due to the absence of available final states.
(C) Schematic energy-level diagram of the KPO, illustrating QCR-induced transitions between eigenstates.
The lowest two levels ($|\phi_0\rangle$, $|\phi_1\rangle$) define the qubit subspace, while higher levels ($|\phi_2\rangle$, $|\phi_3\rangle$) correspond to leakage states.
Two-photon absorption processes preserve the parity of the KPO eigenstates and induce transitions within the same-parity manifold (solid green arrows).
In contrast, single-photon processes couple states of opposite parity (blue and black arrows) and are suppressed in the bias regime shown in (B).
 }
\label{Idea_4_8_26}
\end{figure}

This section describes the physical mechanism of photon-assisted quasiparticle tunneling in a SISIS-based QCR coupled to a KPO.
Figure~\ref{Idea_4_8_26}(A) shows the device structure, in which a superconducting island in a SISIS junction is coupled to the KPO.

When a bias voltage is applied across the junction, quasiparticles can tunnel by absorbing or emitting photons from the KPO.
Figure~\ref{Idea_4_8_26}(B) illustrates photon-assisted quasiparticle tunneling through one of the SIS junctions in the SISIS structure.
The applied bias voltage shifts the quasiparticle energies in one superconductor relative to those in the other.
In the bias regime shown in Fig.~\ref{Idea_4_8_26}(B), a quasiparticle can tunnel by absorbing energy corresponding to two photons of the KPO, thereby bridging the energy difference between the initial and final states.
In contrast, single-photon absorption is forbidden because no available final state exists.
The bias voltage therefore acts as a control parameter that selectively activates different photon-assisted tunneling processes.

The fundamental difference between SINIS- and SISIS-based QCRs lies in how their densities of states shape these tunneling processes.
In SINIS junctions, the normal-metal island provides a smooth density of states, allowing both single- and multi-photon processes over a broad bias range.
As a result, single-photon absorption processes dominate and induce transitions between opposite-parity states of the KPO, leading to phase-flip errors.
In contrast, the sharp features of the superconducting density of states in SISIS junctions can suppress single-photon processes while allowing two-photon absorption processes for appropriately chosen bias voltages, as illustrated in Fig.~\ref{Idea_4_8_26}(B).

As illustrated in Fig.~\ref{Idea_4_8_26}(C), two-photon processes preserve the parity of the KPO eigenstates and therefore induce transitions only within the same-parity manifold.
Consequently, photon-assisted tunneling predominantly couples states of the same parity, while parity-changing transitions, including phase-flip transitions between $|\phi_0\rangle$ and $|\phi_1\rangle$, remain suppressed.

\section{System}
\label{System}
We consider a KPO capacitively coupled to a SISIS junction, as illustrated in Fig.~\ref{Idea_4_8_26}(A). 
The system configuration follows Ref.~\cite{Masuda2025}, except that the normal-metal island is replaced by a superconducting island. 
A bias voltage $V_B$ is applied across the junction, enabling quasiparticle tunneling through the insulating barriers. 
The central superconducting island is capacitively coupled to the KPO, so that charge fluctuations on the island interact with the KPO.

\begin{figure}
\begin{center}
\includegraphics[width=7.5cm]{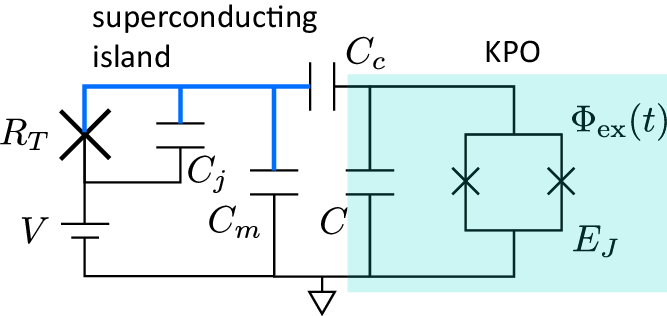}
\end{center}
\caption{
Circuit diagram used in the theoretical analysis of the SISIS-based QCR.
A KPO (shaded area) is capacitively coupled ($C_c$) to a superconducting island (blue lines).
One SIS junction is treated explicitly through the tunnel-junction parameters ($R_T$, $C_j$), while the second SIS junction is modeled as a capacitor whose capacitance is incorporated into the effective island capacitance $C_m$ following Ref.~\cite{Silveri2017}.
The voltage $V$ is the bias voltage across the explicitly treated SIS junction ($V=V_B/2$).
The KPO is modeled by a capacitance $C$ and a SQUID with Josephson energy $E_J$, driven by an external magnetic flux $\Phi_{\mathrm{ex}}(t)$.
}
\label{circuit_KPO_1_11_24}
\end{figure}

Figure~\ref{circuit_KPO_1_11_24} shows the circuit model used in the theoretical analysis.
Following Ref.~\cite{Silveri2017}, quasiparticle tunneling through one SIS junction is treated explicitly, while the capacitance of the second SIS junction is incorporated into the effective island capacitance.
This treatment allows us to describe the effect of photon-assisted quasiparticle tunneling on the KPO while retaining the electrostatic contribution of both SIS junctions.

The tunnel resistance and capacitance of the explicitly treated SIS junction are denoted by $R_T$ and $C_j$, respectively, while $C_m$ represents the effective capacitance of the superconducting island to ground.
The KPO is modeled by a capacitance $C$ and a flux-driven SQUID with Josephson energy $E_J$~\cite{Wang2019}. The parameter $C_c$ denotes the coupling capacitance between the KPO and the superconducting island. A bias voltage $V_B$ is applied across the SISIS structure, corresponding to a voltage drop $V=V_B/2$ across the explicitly treated SIS junction.

\subsection{Hamiltonian}
The dynamics is analyzed in a rotating frame at frequency $\omega_p/2$.
In this frame the total Hamiltonian can be written schematically as
\begin{eqnarray}
H^{(\rm RF)} = H_{\rm env}^{(\rm RF)} + H_{\rm KPO}^{(\rm RF)} + H_T^{(\rm RF)},
\label{H_tot_ver2_8_13_26}
\end{eqnarray}
where $H_{\rm env}^{(\rm RF)}$ describes quasiparticles and the charge degree of freedom of the superconducting island,
$H_{\rm KPO}^{(\rm RF)}$ describes the KPO,
and $H_T^{(\rm RF)}$ describes the coupling between quasiparticle tunneling processes and the KPO.
$H_{\rm env}^{(\rm RF)}$ and $H_T^{(\rm RF)}$ are given in Eq.~(\ref{H_env_8_13_26}) and Eq.~(\ref{H_T_6_1_26}), respectively (see Appendix~\ref{AppendixRates}).

The KPO Hamiltonian is given by
\begin{eqnarray}
H_{\rm KPO}^{\rm (RF)} / \hbar = \Delta_{\rm KPO} a^\dagger a - \frac{\chi}{2} a^\dagger a^\dagger a a
+\beta(a^2 + a^{\dagger 2}),
\label{HKPO_11_15_24}
\end{eqnarray}
where $a$ ($a^\dagger$) is the annihilation (creation) operator of the resonator mode, $\Delta_{\rm KPO}$ is the detuning, $\chi$ is the Kerr nonlinearity, and $\beta$ is the pump amplitude.
See Refs.~\cite{Wang2019,Masuda2025} for details.

The parameters of the KPO are related to the circuit parameters as follows~\cite{Masuda2025}.
The Kerr nonlinearity $\chi$ is given by $\chi=E_C/\hbar$ with $E_C=e^2/(2C_r)$, $C_r = C + \alpha_{c} {C_\Sigma}_m$, $\alpha_{c} =  C_c/C_I$, $C_I = C_c + C_m + C_j$, and ${C_\Sigma}_m = C_m + C_j$. 
The pump amplitude $\beta$ is given by $\beta = \omega_c^{(0)}\delta E_J/(8E_J)$, where $\omega_c^{(0)}=\frac{1}{\hbar}\sqrt{8E_CE_J}$.
The detuning is given by $\Delta_{\rm KPO}=\omega_c-\omega_p/2$, where $\omega_c$ is the dressed resonator frequency defined by $\omega_c=\omega_c^{(0)}-\chi$.
We assume that the Josephson energy $E_J(t)$ is modulated  as $E_J(t)=E_J+\delta E_J \cos(\omega_p t)$ through a time-dependent magnetic flux applied to the SQUID.

In this work, we focus on the case $\Delta_{\rm KPO}=0$, where the lowest two eigenstates of the isolated KPO form a degenerate qubit subspace. 
These states are given by
\begin{eqnarray}
|\phi_0\rangle &=& N_+ (|\alpha\rangle + |-\alpha\rangle), \nonumber\\
|\phi_1\rangle &=& N_- (|\alpha\rangle - |-\alpha\rangle),
\end{eqnarray}
where $|\pm\alpha\rangle$ are coherent states with $\alpha=\sqrt{2\beta/\chi}$ and $N_{\pm} = (2\pm 2e^{-2\alpha^2})^{-1/2}$. 
Higher excited states constitute leakage states outside the computational subspace, as illustrated in Fig.~\ref{Idea_4_8_26}(C).

Because the Hamiltonian conserves parity, the eigenstates can be classified into even- and odd-parity manifolds, $|\phi_{2n}\rangle$ (even) and $|\phi_{2n+1}\rangle$ (odd).

The coupling between the KPO and the superconducting island is characterized by the dimensionless parameter~\cite{Silveri2017}
\begin{eqnarray}
\rho_c=\alpha_c^2\sqrt{\frac{E_C}{8E_J}}.
\end{eqnarray}
The parameter $\rho_c$ determines the strength of photon-assisted quasiparticle tunneling induced by the capacitive coupling between the KPO and the superconducting island.

\subsection{Effective QCR-induced dynamics}
\label{Effective QCR-induced dynamics}
In this section, we present an effective equation of motion for the KPO that incorporates photon-assisted quasiparticle tunneling.
Our derivation closely follows Ref.~\cite{Masuda2025}, except that the normal-metal island is replaced by a superconducting island, which modifies the quasiparticle density of states and hence the tunneling rates.

The effect of quasiparticle tunneling is treated perturbatively with respect to the tunneling Hamiltonian $H_T$.
The total system is divided into the KPO and its environment, where the environment consists of quasiparticles in the superconducting electrodes and the charge degree of freedom of the superconducting island.

The reduced density-matrix elements in the eigenbasis $\{|\phi_\mu\rangle\}$ of the KPO Hamiltonian are defined as
\begin{eqnarray}
\rho^{\rm KPO}_{\phi_{\mu}, \phi_{\mu'}}(t) 
= \sum_{\mathcal{E}_\mu} \langle \phi_{\mu}, \mathcal{E}_\mu | \rho_{\rm tot}(t) | \phi_{\mu'}, \mathcal{E}_\mu \rangle,
\label{rho_sys_2_14_24}
\end{eqnarray} 
where $\rho_{\rm tot}$ is the density operator of the total system, and $|\mathcal{E}_\mu\rangle$ denotes a joint state of the superconducting island and quasiparticles.

Tracing out the quasiparticles and the superconducting island within a Born--Markov approximation in the rotating frame, we obtain an effective equation of motion for the reduced density matrix of the KPO in its eigenbasis $\{|\phi_\mu\rangle\}$~\cite{Masuda2025}:
\begin{eqnarray}
&&\frac{d}{dt}
\rho_{\phi_{\mu},\phi_{\mu'}}^{\rm KPO}(t) =
-i\omega_{\phi_{\mu},\phi_{\mu'}}
\rho_{\phi_{\mu},\phi_{\mu'}}^{\rm KPO}(t)
\nonumber\\
&&\hspace{1cm} +
\sum_{\phi_\nu}\sum_{\phi_{\nu'}}'
\Gamma^{(1)}
(\phi_{\mu},\phi_{\mu'},\phi_\nu,\phi_{\nu'},V)
\rho_{\phi_\nu,\phi_{\nu'}}^{\rm KPO}(t)
\nonumber\\
&&\hspace{1cm}
+ \sum_{\phi_{\xi}}' \Gamma^{(2)} (\phi_{\mu},\phi_{\mu'},\phi_{\xi},V)
\rho_{\phi_\xi,\phi_{\mu'}}^{\rm KPO}(t)
\nonumber\\
&&\hspace{1cm}
+ \sum_{\phi_{\xi}}'' \Gamma^{(3)} (\phi_{\mu},\phi_{\mu'},\phi_{\xi},V)
\rho_{\phi_{\mu},\phi_{\xi}}^{\rm KPO}(t),
\label{EOM_2_26_26}
\end{eqnarray}
where $\omega_{\phi_{\mu},\phi_{\mu'}} =(E_{\phi_{\mu}} - E_{\phi_{\mu'}})/\hbar$.
The coefficients $\Gamma^{(1)}$, $\Gamma^{(2)}$, and $\Gamma^{(3)}$ are obtained from a Born--Markov treatment of quasiparticle tunneling based on Fermi's golden rule.
Since they originate from tunneling processes, all coefficients are proportional to $1/R_T$, where $R_T$ is the tunnel resistance.
Their explicit forms and derivation are given in Appendix~\ref{AppendixRates}.

From Eq.~(\ref{EOM_2_26_26}), the diagonal elements
\begin{eqnarray}
\Gamma_{j \rightarrow i}(V)
\equiv
\Gamma^{(1)}(\phi_i,\phi_i,\phi_j,\phi_j,V)
\label{Gamma_6_18_26}
\end{eqnarray}
represent QCR-induced transition rates between the KPO eigenstates $|\phi_i\rangle$ and $|\phi_j\rangle$.
The remaining coefficients, $\Gamma^{(2)}$ and $\Gamma^{(3)}$, are required for an accurate description of the time evolution of the KPO density matrix.

Finally, we note that the present derivation relies on the rotating-wave approximation and a Born--Markov treatment of quasiparticle tunneling.
We focus on the regime where the applied bias voltage is comparable to the superconducting gap, such that quasiparticle tunneling dominates the transport.
The complementary low-bias regime, where Cooper-pair tunneling becomes important~\cite{Chen2014,Cassidy2017}, lies beyond the scope of this work.
Accordingly, Cooper-pair tunneling processes are neglected throughout this work, and quasiparticle tunneling provides the dominant mechanism for dissipation and energy exchange between the SISIS junction and the KPO.

\section{Results}
\label{Results}
\begin{figure}[]
\centering
\includegraphics[width=8.5cm]{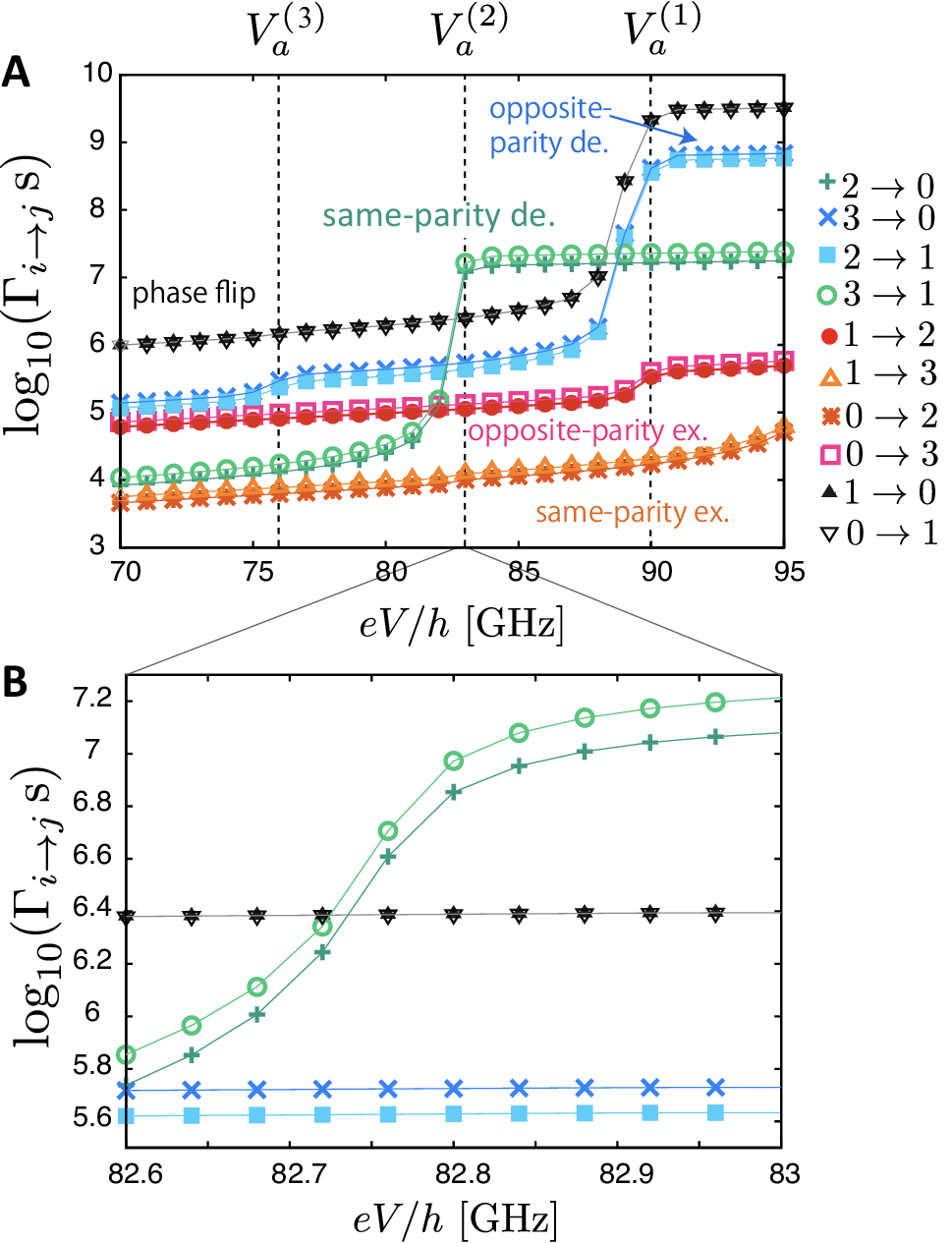}
\caption{
Transition rates $\Gamma_{i\rightarrow j}$
[Eq.~(\ref{Gamma_6_18_26})] as functions of the normalized bias voltage
$eV/h$ for $\alpha=2$ and $\rho_c=0.01$.
Panel (A) shows the rates over a wide voltage range including the thresholds
$V_a^{(n)}$ ($n=1,2,3$), while panel (B) presents an enlarged view around
$V_a^{(2)}$, highlighting the deexcitation and phase-flip rates.
Vertical dashed lines in panel (A) indicate the threshold voltages
$V_a^{(n)}$ associated with the onset of one-, two-, and three-photon
absorption processes.
The parameters are
$\chi/2\pi=10$~MHz,
$\omega_c/2\pi=7$~GHz,
$\Delta=200~\mu$eV,
$\Delta_{\rm KPO}/2\pi=0$,
$R_{\rm T}=50$~k$\Omega$,
$\gamma_{\rm D}=10^{-4}$,
$\beta/2\pi=20$~MHz,
and $T_{S}=100$~mK,
which are comparable to experimental values~\cite{Yamaji2022,Tan2017}.
}
\label{Gamma1_V_4_2_26}
\end{figure}

We investigate the bias-voltage dependence of QCR-induced transition rates in the KPO.
Figure~\ref{Gamma1_V_4_2_26} shows the dominant transitions relevant to cooling, heating, and phase-flip processes.
The rates are calculated using the formalism developed in Sec.~\ref{Effective QCR-induced dynamics} and Appendix~\ref{AppendixRates}.
To isolate the effect of QCR-induced transitions, other decoherence mechanisms, such as single-photon loss and pure dephasing, are neglected unless otherwise stated.

As the bias voltage increases, the deexcitation rates between same-parity states 
($\phi_2\rightarrow\phi_0$ and $\phi_3\rightarrow\phi_1$)
exhibit pronounced increases around the two-photon absorption threshold
$V_a^{(2)}=(2\Delta-2\hbar\omega_{\rm RF})/e$,
indicating deexcitation mediated by the absorption of two photons from the KPO.
In contrast, the deexcitation rates between opposite-parity states
($\phi_3\rightarrow\phi_0$ and $\phi_2\rightarrow\phi_1$)
become appreciable already around the single-photon absorption threshold
$V_a^{(1)}=(2\Delta-\hbar\omega_{\rm RF})/e$,
reflecting single-photon absorption processes, which are also responsible for QCR-induced phase-flip errors.

A particularly important feature is that the phase-flip rate,
corresponding to transitions between $\phi_0$ and $\phi_1$,
is strongly enhanced near $V_a^{(1)}$ but exhibits no comparable enhancement near $V_a^{(2)}$.
This behavior is consistent with the fact that QCR-induced phase flips are predominantly associated with single-photon absorption processes~\cite{Masuda2025},
whereas two-photon absorption mainly induces transitions within the same-parity manifold.
Therefore, operation in the bias regime where two-photon absorption dominates allows the KPO to be cooled while minimizing QCR-induced phase-flip errors.
In the following, we fix the bias voltage near this operating point $V_a^{(2)}$ and investigate the dependence of the transition rates on other parameters.

\begin{figure}[]
\centering
\includegraphics[width=8.5cm]{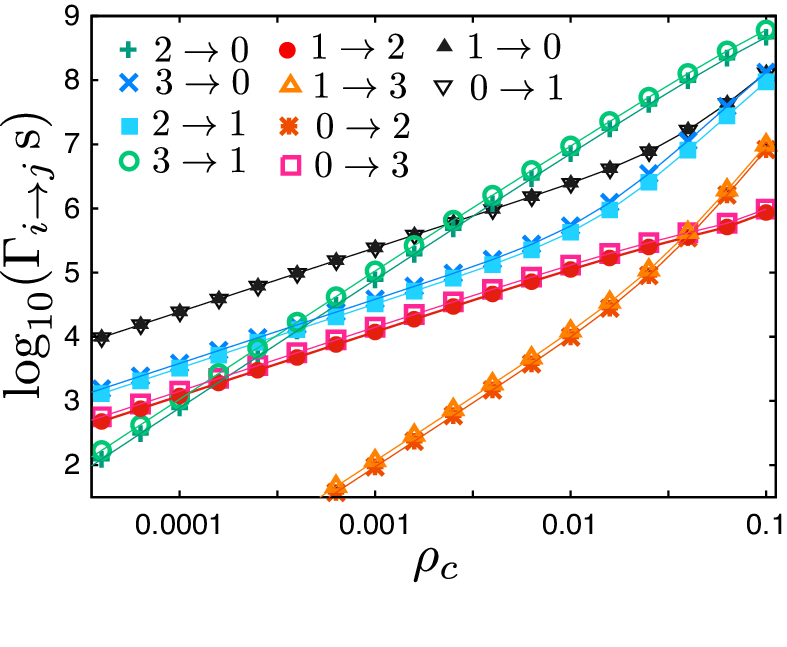}
\caption{
$\rho_c$ dependence of QCR-induced transition rates between KPO eigenstates at $eV/h=82.8$~GHz ($\simeq eV_a^{(2)}/h$) for $\alpha=2$.
The color scheme and other parameters are the same as in Fig.~\ref{Gamma1_V_4_2_26}. 
}
\label{Gamma1_rho_6_18_26}
\end{figure}

Figure~\ref{Gamma1_rho_6_18_26} shows the dependence of the QCR-induced transition rates on $\rho_c$, which characterizes the coupling strength between the KPO and the superconducting island.
It is seen that the same-parity deexcitation rates dominate both the excitation rates and the phase-flip rates for $\rho_c\gtrsim0.01$, indicating efficient cooling in this regime.
However, increasing $\rho_c$ does not always improve the cooling performance.
For sufficiently large $\rho_c$, the same-parity excitation rates grow more rapidly than the corresponding deexcitation rates, leading to enhanced unwanted excitation processes.

\begin{figure}
\begin{center}
\includegraphics[width=7.5cm]{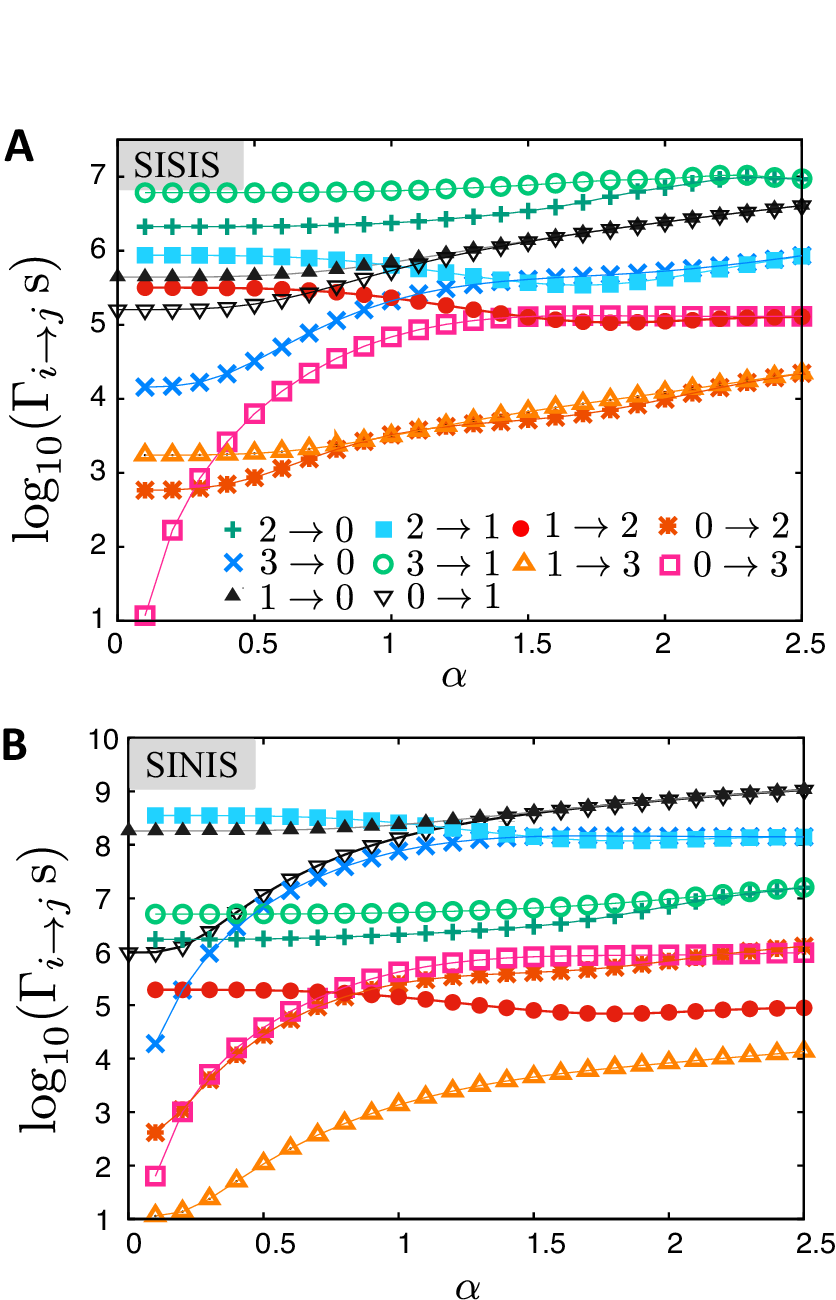}
\end{center}
\caption{
$\alpha$ dependence of relevant transition rates for a SISIS junction at
$eV/h=82.8$~GHz ($\simeq eV_a^{(2)}/h$) with $\rho_c=0.01$
(A), in the two-photon-dominant regime, and for a SINIS junction at
$eV/h=45$~GHz ($\simeq eV_a^{(1)}/h$) with $\rho_c=0.01$
(B), in the single-photon-dominant regime.
The color scheme and other parameters are the same as in Fig.~\ref{Gamma1_V_4_2_26}. 
}
\label{Gamma1_alpha_4_3_26}
\end{figure}

Figure~\ref{Gamma1_alpha_4_3_26} shows the $\alpha$ dependence of relevant transition rates for SISIS and SINIS junctions.
For SISIS, the transition rates are evaluated at the same operating point as in Fig.~\ref{Gamma1_rho_6_18_26}.
For SINIS, the bias voltage is chosen such that single-photon absorption is dominant ($eV/h=45$~GHz).
The coupling parameter is fixed at $\rho_c=0.01$.

In both configurations, the deexcitation rates change only moderately with $\alpha$ for sufficiently large $\alpha$, indicating that the cooling performance is robust against variations in the coherent-state amplitude.
Thus, the robustness of the cooling performance is not unique to the SISIS-based QCR.

The key difference appears in the phase-flip rate.
For SISIS, the phase-flip rate remains lower than the dominant cooling rates,
$\Gamma_{2\rightarrow0}$ and $\Gamma_{3\rightarrow1}$,
over the range of $\alpha$ considered here,
whereas for SINIS it increases substantially at larger $\alpha$.
These results demonstrate that the SISIS-based QCR maintains efficient cooling while suppressing phase-flip errors compared with the SINIS-based approach.

\begin{figure}
\includegraphics[width=7.5cm]{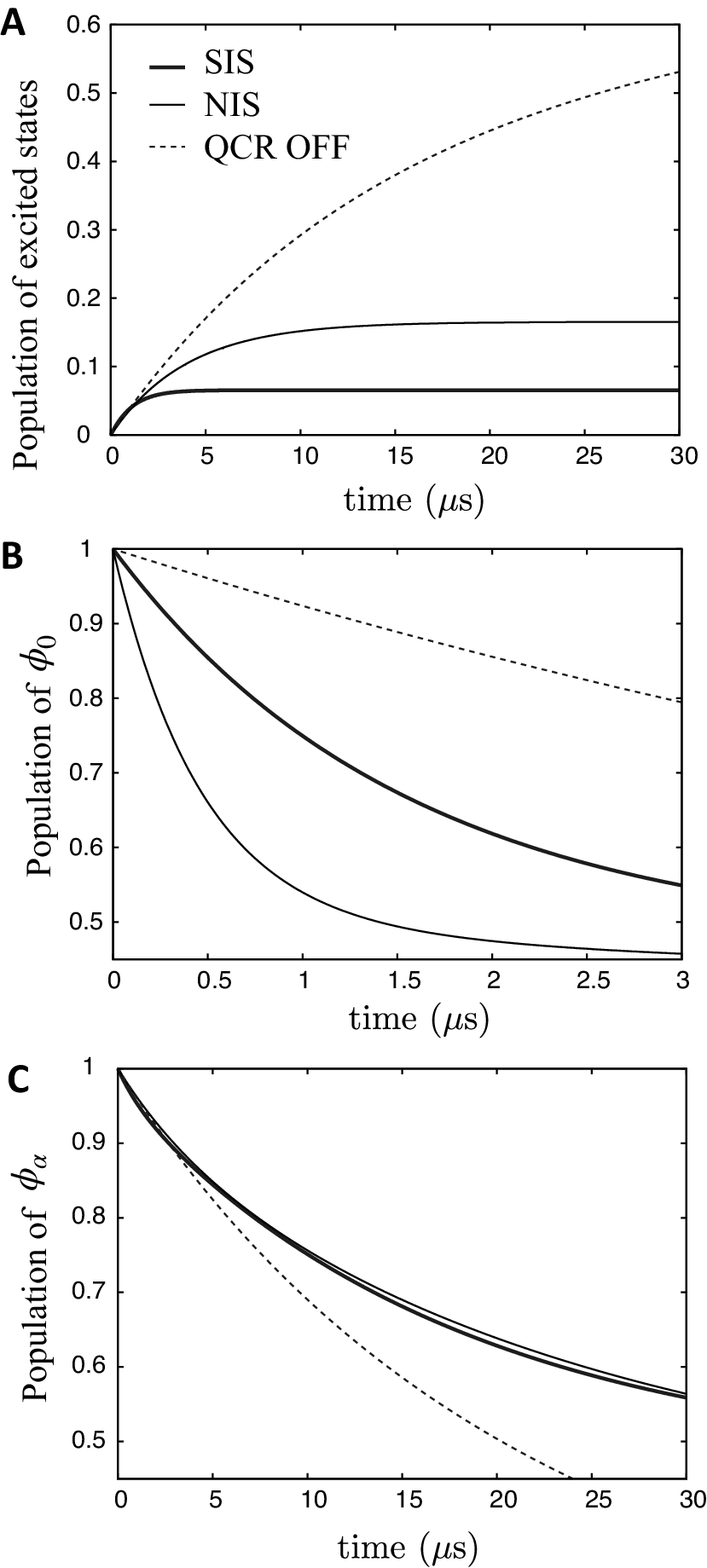}
\caption{
Time evolution of state populations under pure dephasing, internal loss, and QCR. 
(A) Population of excited (leakage) states as a function of time for SISIS (thick solid line), SINIS (thin solid line), and QCR off (dotted line).  
(B) Population of $|\phi_0\rangle$. 
A shorter time window is shown to highlight the effect of QCR-induced phase-flip processes.
(C) Population of $|\phi_\alpha\rangle$. 
The initial state is $|\phi_0\rangle$ in (A) and (B), and $|\phi_\alpha\rangle$ in (C).
The values of the internal loss rate $\kappa$ and pure dephasing rate $\gamma_p$ are
$\kappa/2\pi=1.6$~kHz and $\gamma_p/2\pi=0.8$~kHz, respectively,
and are comparable to the ones measured for flux-tunable superconducting qubits~\cite{Chavez-Garcia2022,Tuokkola2024}.
The tunnel resistance is $R_T=500$~k$\Omega$ for SISIS and
$R_T=50$~k$\Omega$ for SINIS.
The bias voltage is chosen such that two-photon absorption is dominant for SISIS
($eV/h=82.8$~GHz) and single-photon absorption is dominant for SINIS
($eV/h=40$~GHz).
The coupling parameters are $\rho_c=0.01$ for SISIS and
$\rho_c=5\times10^{-5}$ for SINIS.
All remaining parameters for SISIS are as in Fig.~\ref{Gamma1_V_4_2_26};
the SINIS parameters are identical to those used in Ref.~\cite{Masuda2025}.
}
\label{fig_time_deps_4_13_26}
\end{figure}

We numerically simulate the dynamics of the KPO under the combined effects of pure dephasing, internal loss, and the QCR (see Appendix~\ref{Details of time-evolution calculations} for details).
Figure~\ref{fig_time_deps_4_13_26} shows the time evolution of the populations of relevant states.
The initial state is set to $|\phi_0\rangle$ in panels (A) and (B), and to $|\phi_\alpha\rangle$ in panel (C).
Here,
$|\phi_{\pm\alpha}\rangle=(|\phi_0\rangle \pm |\phi_1\rangle)/\sqrt{2}$,
which satisfy
$|\phi_{\pm\alpha}\rangle \simeq |\pm\alpha\rangle$
for sufficiently large $\alpha$.
We compare SISIS- and SINIS-based QCRs, with the bias voltage chosen such that two-photon absorption is dominant for SISIS and single-photon absorption is dominant for SINIS.
The tunnel resistance is chosen to be larger for SISIS than for SINIS. 
The purpose of this comparison is not to maximize the cooling power of either configuration, but to examine whether leakage can be suppressed with a smaller phase-flip penalty.

Before discussing the results, we emphasize that the key issue is not the cooling power itself.
A SINIS-based QCR can also achieve strong cooling by appropriately tuning its parameters.
However, stronger cooling in the SINIS configuration is generally accompanied by an increase in QCR-induced phase-flip errors because the cooling mechanism relies on single-photon absorption.
The central question is therefore whether a SISIS-based QCR can suppress leakage while avoiding this penalty through two-photon-assisted cooling.

As shown in Fig.~\ref{fig_time_deps_4_13_26}(A), the SISIS-based QCR suppresses the leakage population,
$P_{\rm leak}=\sum_{\mu\ge2}\rho^{\rm KPO}_{\phi_\mu,\phi_\mu}$,
more effectively than the SINIS-based QCR.
Importantly, Fig.~\ref{fig_time_deps_4_13_26}(B) shows that the population of $|\phi_0\rangle$ remains higher for SISIS than for SINIS.
Together, these results indicate that SISIS suppresses leakage more efficiently while incurring a smaller phase-flip penalty.
Although a SINIS-based QCR can achieve stronger cooling by increasing the cooling power, this generally comes at the cost of more frequent phase-flip errors.
The SISIS-based QCR alleviates this trade-off by exploiting two-photon-assisted cooling processes, which suppress parity-changing transitions associated with phase-flip errors.
Although the population of $|\phi_0\rangle$ is highest when the QCR is turned off, the associated leakage population is also significantly larger.
Since leakage errors cannot be corrected by conventional quantum-error-correction schemes, reducing leakage is essential for preserving the encoded quantum information.

Figure~\ref{fig_time_deps_4_13_26}(C) shows the population of $|\phi_{\pm\alpha}\rangle$.
The decay of $|\phi_{\pm\alpha}\rangle$ is comparable for SISIS and SINIS, indicating similar bit-flip rates in the two configurations.
Furthermore, the lifetime of $|\phi_{\pm\alpha}\rangle$ is much longer than that of $|\phi_0\rangle$, reflecting the biased-noise nature of Kerr-cat qubits, where bit-flip errors are strongly suppressed relative to phase-flip errors.
When the QCR is turned off, $|\phi_{\pm\alpha}\rangle$ decays more rapidly because leakage to excited states is no longer reduced by QCR-assisted relaxation.

Since the QCR-induced transition rates scale inversely with the tunnel resistance [Eq.~(\ref{Gamma_2_27_27})], reducing $R_T$ enhances both the desired cooling transitions and unwanted phase-flip processes.
Therefore, although the SISIS-based QCR suppresses phase-flip errors more effectively than the SINIS-based QCR, excessively strong cooling can still degrade performance through an increase in the phase-flip rate (see Appendix~\ref{Dependence on tunnel resistance}).
This behavior reflects a residual cooling--phase-flip trade-off that remains even in the SISIS configuration.
Consequently, the QCR operating parameters should be chosen to provide a favorable balance between leakage suppression and phase-flip errors.

\section{Conclusion}
In this work, we have theoretically investigated quantum circuit refrigeration (QCR) based on a SISIS junction coupled to a Kerr parametric oscillator (KPO), and compared its performance with the conventional SINIS-based QCR. 
A key feature of the SISIS-based QCR is the existence of an operating regime in which cooling is mediated predominantly by two-photon absorption processes.

By analyzing photon-assisted quasiparticle tunneling, we have shown that the SISIS-based QCR enables efficient cooling of the KPO via two-photon absorption, while strongly suppressing QCR-induced phase-flip errors. 
This behavior can be understood as an energy-filtering effect arising from the sharp superconducting density of states, which selectively suppresses single-photon processes while favoring two-photon absorption processes.
In contrast, the SINIS-based QCR, which relies on single-photon absorption, inevitably enhances phase-flip errors when operated at high cooling power.

Our results demonstrate that the SISIS-based QCR can achieve effective cooling power and robust stabilization of the qubit subspace, with phase-flip rates significantly lower than those of the SINIS-based approach.
Importantly, the bit-flip rates remain comparable between the two schemes, indicating that the main advantage of the SISIS-based QCR lies in its ability to suppress phase-flip errors without sacrificing cooling efficiency.

We have also shown that the cooling performance and error bias of the SISIS-based QCR are robust against moderate parameter variations, such as the coherent-state amplitude $\alpha$.
Although excessive cooling power can increase phase-flip errors even in the SISIS configuration, an intermediate operating regime provides a favorable balance between leakage suppression and preservation of the biased error structure.

These findings highlight the potential of SISIS-based QCRs as an effective approach for mitigating leakage while limiting QCR-induced phase-flip errors in Kerr-cat qubits.

\section*{ACKNOWLEDGEMENTS}
SM acknowledges support from the JST Moonshot R\&D-MILLENNIA program (Grant No. JPMJMS2061).
SN thanks the support from JSPS KAKENHI Grant Number JP25K01611and JST ASPIRE Grant Number JPMJAP25A2.
TY was supported by JSPS Overseas Research Fellowships.

\appendix

\section{Derivation of QCR-induced transition rates}
\label{AppendixRates}
In this Appendix, we summarize the derivation of the effective equation of motion used in the main text. 
The procedure follows Ref.~\cite{Masuda2025}, with the essential modification that both electrodes are superconducting, so that the superconducting density of states appears for both sides of the SISIS junction.

\subsection{Microscopic Hamiltonian}
In the lab frame, the total system, consisting of quasiparticles, the superconducting island, and the KPO, is described by
\begin{eqnarray}
H_{\rm tot} = H_{\rm QP} + H_T + H_0,
\label{Htot_6_18_26}
\end{eqnarray}
where $H_{\rm QP}$ describes quasiparticles in the superconducting electrodes, $H_T$ is the tunneling Hamiltonian, and $H_0$ describes the coupled KPO--island circuit.

Here, $H_{\rm QP}$ is written as
\begin{eqnarray}
H_{\rm QP} = \sum_{k,\sigma} (\varepsilon_k - eV) c_{k\sigma}^\dagger c_{k\sigma}
+ \sum_{l,\sigma} \varepsilon_l d_{l\sigma}^\dagger d_{l\sigma},
\label{Hqp_1_30_23}
\end{eqnarray}
where $c_{k\sigma}$ ($d_{l\sigma}$) annihilates a quasiparticle with spin $\sigma$ in the right (left) superconducting electrode, and $\varepsilon_k$ and $\varepsilon_l$ are the corresponding quasiparticle energies. 
The energy shift $-eV$ accounts for the applied bias voltage $V=V_B/2$.

The tunneling Hamiltonian $H_T$ describes quasiparticle tunneling across the junction and its coupling to the superconducting circuit,
\begin{eqnarray}
H_T = \sum_{k,l,\sigma} T_{lk} d_{l\sigma}^\dagger c_{k\sigma} e^{-i\varphi_I} + \mathrm{h.c.},
\label{HT_2_27_27}
\end{eqnarray}
where $\varphi_I$ is the dimensionless flux associated with the superconducting island. 
The operator $e^{-i\varphi_I}$ shifts the island charge by one elementary charge, corresponding to a quasiparticle tunneling event.

The Hamiltonian of the effective circuit is given by
\begin{eqnarray}
H_0 &=& \frac{(Q_I + Q_j)^2}{2C_I}
+ \frac{\big[ Q + \alpha_c (Q_I + Q_j) \big]^2}{2C_r} \nonumber\\
&& - E_J(t) \cos\!\left( \frac{2e}{\hbar}\Phi \right),
\end{eqnarray}
where $\Phi$ and $Q$ are the flux and charge of the KPO, satisfying $[\Phi,Q]=i\hbar$, and $Q_I$ is the charge of the superconducting island conjugate to $\Phi_I$. 
The Josephson energy is parametrically modulated as $E_J(t)=E_J+\delta E_J \cos(\omega_p t)$.

\subsection{Transformation to the rotating frame}
We use the composite unitary transformation
\begin{eqnarray}
U_{\rm tot}(t) = U_{\rm RF}(t)\, U\, U_j,
\end{eqnarray}
where
$U_j = \exp\!\left( \frac{i}{\hbar} Q_j \Phi_I \right)$ eliminates the offset charge of the superconducting island,
$U = \sum_q \exp\!\left( \frac{i}{\hbar} \alpha_c e q \Phi \right) |q\rangle\langle q|$ removes the coupling-induced shift of the KPO charge,
and
$U_{\rm RF}(t) = \exp\!\left( i \frac{\omega_p}{2} t\, a^\dagger a \right)$ moves the system into a rotating frame at frequency $\omega_p/2$.

Within the rotating-wave approximation, the transformed Hamiltonian is written as
\begin{eqnarray}
H^{\rm (RF)} = H_{\rm QP} + H_0^{\rm (RF)} + H_T^{\rm (RF)} ,
\label{H_tot_ver1_8_13_26}
\end{eqnarray}
where $H_{\rm QP}$ is unchanged by the transformation.
The circuit Hamiltonian takes the form
\begin{eqnarray}
H_0^{\rm (RF)} = \sum_q \frac{e^2 q^2}{2C_I} |q\rangle\langle q| + H_{\rm KPO}^{\rm (RF)},
\label{H0_RF_12_19_23}
\end{eqnarray}
with $Q_I |q\rangle = e q |q\rangle$.
Here, $H_{\rm KPO}^{\rm (RF)}$ is the effective Hamiltonian of the KPO in the rotating frame, given in Eq.~(\ref{HKPO_11_15_24}).
Equation~(\ref{H_tot_ver2_8_13_26}) is obtained by introducing the environmental Hamiltonian
\begin{eqnarray}
H_{\rm env}^{(\rm RF)} = H_{\rm QP} + \sum_q \frac{e^2 q^2}{2C_I} |q\rangle\langle q|,
\label{H_env_8_13_26}
\end{eqnarray}
where the environment consists of quasiparticles in the superconducting electrodes and the charging degree of freedom of the superconducting island.

The transformed tunneling Hamiltonian $H_T^{\rm (RF)}(t)$ couples quasiparticle tunneling to transitions between KPO eigenstates and is treated perturbatively in the following.
It is given by
\begin{eqnarray}
H_T^{\rm (RF)}(t) &=& 
\sum_{m,m'} \sum_{k,l,\sigma} \sum_q 
e^{ i \omega_{\rm RF} (m'-m) t } \nonumber\\
&& \times
\Big[
\langle m' | e^{-\frac{i}{\hbar}\alpha_c e \Phi} | m \rangle\,
T_{lk}\, d_{l\sigma}^\dagger c_{k\sigma}\,
|q-1,m'\rangle\langle q,m| \nonumber\\
&&  +
\langle m' | e^{\frac{i}{\hbar}\alpha_c e \Phi} | m \rangle\,
T_{lk}^\ast\, c_{k\sigma}^\dagger d_{l\sigma}\,
|q+1,m'\rangle\langle q,m|
\Big], \nonumber\\
\label{H_T_6_1_26}
\end{eqnarray}
where $|q,m\rangle = |q\rangle\otimes|m\rangle$ denotes a product state of the island charge and the KPO Fock state.
This form explicitly shows that quasiparticle tunneling events are accompanied by the absorption or emission of an integer number of photons of the KPO.

\subsection{Born--Markov treatment}
The tunneling Hamiltonian $H_T^{\rm (RF)}$ is treated perturbatively to second order.
Starting from the von Neumann equation for the total density matrix and applying the Born--Markov approximation, we trace out the quasiparticle and superconducting-island degrees of freedom.
This procedure yields a coarse-grained equation of motion for the reduced density matrix of the KPO in the eigenbasis $\{|\phi_\mu\rangle\}$:
\begin{eqnarray}
&&\rho_{\phi_{\mu},\phi_{\mu'}}^{\rm KPO}(t+\Delta t) = \rho_{\phi_{\mu},\phi_{\mu'}}^{\rm KPO}(t)
- i\omega_{\phi_{\mu},\phi_{\mu'}} \Delta t \rho_{\phi_{\mu},\phi_{\mu'}}^{\rm KPO}(t) \nonumber\\
&& \hspace{1cm} + \sum_{\phi_\nu}\sum_{\phi_{\nu'}}' \Gamma^{(1)}(\phi_{\mu},\phi_{\mu'},\phi_\nu,\phi_{\nu'},V) \Delta t
\rho_{\phi_\nu,\phi_{\nu'}}^{\rm KPO}(t) \nonumber\\
&& \hspace{1cm} + \sum_{\phi_{\xi}}' \Gamma^{(2)}(\phi_{\mu},\phi_{\mu'},\phi_{\xi},V) \Delta t
\rho_{\phi_\xi,\phi_{\mu'}}^{\rm KPO}(t)\nonumber\\
&& \hspace{1cm} + \sum_{\phi_{\xi}}'' \Gamma^{(3)}(\phi_{\mu},\phi_{\mu'},\phi_{\xi},V) \Delta t
\rho_{\phi_{\mu},\phi_{\xi}}^{\rm KPO}(t).
\end{eqnarray}
In the limit $\Delta t\rightarrow0$, this equation reduces to the differential form given in Eq.~(\ref{EOM_2_26_26}) of the main text. 
The coefficients $\Gamma^{(1)}$, $\Gamma^{(2)}$, and $\Gamma^{(3)}$ describe QCR-induced population transfer and decoherence processes.

\subsection{Explicit expressions for the rates}
The coefficients are obtained from Fermi's golden rule applied to photon-assisted quasiparticle tunneling processes.
$\Gamma^{(i)}$ in Eq.~(\ref{EOM_2_26_26}) are defined by
\begin{eqnarray}
&& \Gamma^{(1)}(\phi_{\mu},\phi_{\mu'},\phi_\nu,\phi_{\nu'},V) =
\frac{2}{e^2{R_T}} \sum_{\delta m,\delta m',q}  p_{q}  \nonumber\\
&& \times \Big{[} \int d\varepsilon_k n_s(\varepsilon_k) n_s(\varepsilon_l^{(f,\delta m,1)})\nonumber\\
&& \times
[1-f(\varepsilon_k,T_R)] f(\varepsilon_l^{(f,\delta m,1)},T_L)
\eta_{\phi_{\mu},\phi_{\nu}}^{(f,\delta m)} \big{(}\eta_{\phi_{\mu'},\phi_{\nu'}}^{(f,\delta m')}\big{)}^\ast 
\nonumber\\
&& + \int d\varepsilon_k n_s(\varepsilon_k) n_s(\varepsilon_l^{(b,\delta m,1)})\nonumber\\
&& \times
f(\varepsilon_k,T_R) [1-f(\varepsilon_l^{(b,\delta m,1)},T_L)]
\eta_{\phi_{\mu},\phi_{\nu}}^{(b,\delta m)} (\eta_{\phi_{\mu'},\phi_{\nu'}}^{(b,\delta m')})^\ast \Big{]},
\nonumber\\
&& \Gamma^{(2)}(\phi_{\mu},\phi_{\mu'},\phi_\xi,V) =
-\frac{1}{e^2{R_T}} \sum_{\delta m,\delta m',q} \sum_{\phi_\nu} p_{q}  \nonumber\\
&& \times \Big{[} \int d\varepsilon_k n_s(\varepsilon_k) n_s(\varepsilon_l^{(f,\delta m,2)})\nonumber\\
&& \times
[1-f(\varepsilon_k,T_R)] f(\varepsilon_l^{(f,\delta m, 2)},T_L)
\big{(}\eta_{\phi_{\nu},\phi_{\mu}}^{(f,\delta m)}\big{)}^\ast \eta_{\phi_{\nu},\phi_{\xi}}^{(f,\delta m')} 
\nonumber\\
&& + \int d\varepsilon_k n_s(\varepsilon_k) n_s(\varepsilon_l^{(b,\delta m,2)})\nonumber\\
&& \times
f(\varepsilon_k,T_R) [1-f(\varepsilon_l^{(b,\delta m,2)},T_L)]
\big{(} \eta_{\phi_{\nu},\phi_{\mu}}^{(b,\delta m)} \big{)}^\ast \eta_{\phi_{\nu},\phi_{\xi}}^{(b,\delta m')} \Big{]},
 \nonumber\\
&& \Gamma^{(3)}(\phi_{\mu},\phi_{\mu'},\phi_\xi,V) =
-\frac{1}{e^2{R_T}} \sum_{\delta m,\delta m',q} \sum_{\phi_\nu} p_{q}  \nonumber\\
&& \times \Big{[} \int d\varepsilon_k n_s(\varepsilon_k) n_s(\varepsilon_l^{(f,\delta m,3)})\nonumber\\
&& \times
[1-f(\varepsilon_k,T_R)] f(\varepsilon_l^{(f,\delta m,3)},T_L)
\eta_{\phi_\nu,\phi_{\mu'}}^{(f,\delta m)} \big{(}\eta_{\phi_{\nu},\phi_{\xi}}^{(f,\delta m')}\big{)}^\ast  
\nonumber\\
&& + \int d\varepsilon_k n_s(\varepsilon_k) n_s(\varepsilon_l^{(b,\delta m,3)})\nonumber\\
&& \times
f(\varepsilon_k,T_R) [1-f(\varepsilon_l^{(b,\delta m,3)},T_L)]
\eta_{\phi_\nu,\phi_{\mu'}}^{(b,\delta m)} \big{(}\eta_{\phi_{\nu},\phi_{\xi}}^{(b,\delta m')}\big{)}^\ast  \Big{]},
 \nonumber\\
 \label{Gamma_2_27_27}
\end{eqnarray} 
where $n_s$ is the density of states of the quasiparticles in the superconducting electrode given by
\begin{eqnarray}
n_s(\varepsilon) = \Big{|}{\rm Re} \Big\{ \frac{\varepsilon + i\gamma_{\rm D}\Delta}{\sqrt{(\varepsilon + i\gamma_{\rm D}\Delta)^2 - \Delta^2}} \Big\} \Big{|},
\end{eqnarray} 
with the superconductor gap parameter $\Delta$ and the Dynes parameter $\gamma_{\rm D}$~\cite{Dynes1978}.

Importantly, in contrast to our previous work~\cite{Masuda2025}, the rates $\Gamma^{(i)}$ contain the superconducting density of states $n_s(\varepsilon_l)$ for both electrodes, reflecting the SISIS junction configuration. This qualitative modification of the tunneling spectrum is the key ingredient enabling the dominance of multi-photon absorption processes discussed in this work.

The Fermi-Dirac distribution function is defined by $f(E,T) = 1/[e^{E/(k_B T)}+1]$ with $k_B$ the Boltzmann constant, and $T_L$ and $T_R$ the electron temperature at the left and right  superconducting electrodes, respectively.
The probability, denoted by $p_q$, that the state of the superconducting island is $|q\rangle$, is determined using the elastic tunneling of quasiparticles, in which quasiparticles do not exchange energy with the KPO (see section ``Probability $p_q$" in the following section).

In Eq.~(\ref{EOM_2_26_26}), $\sum_{\phi_{\nu'}}'$, $\sum_{\phi_\xi}'$, and $\sum_{\phi_\xi}''$, respectively, denote the summation with respect to the state of the KPO, $\phi_{\nu'}$, $\phi_{\xi}$, and $\phi_{\xi}$, which satisfy
\begin{eqnarray}
E_{\phi_{\mu}} - E_{\phi_{\nu}} + \frac{\hbar\omega_p\delta m}{2} &=& E_{\phi_{\mu'}} - E_{\phi_{\nu'}} + \frac{\hbar\omega_p\delta m'}{2}, \nonumber\\
-E_{\phi_{\mu}}  + \frac{\hbar\omega_p\delta m}{2} &=& -E_{\phi_{\xi}} + \frac{\hbar\omega_p\delta m'}{2},\nonumber\\
-E_{\phi_{\mu'}}  + \frac{\hbar\omega_p\delta m}{2} &=& -E_{\phi_{\xi}} + \frac{\hbar\omega_p\delta m'}{2}.
\label{matching_2_27_26}
\end{eqnarray}

$\eta_{\phi_{\mu},\phi_{\nu}}^{(f,\delta m)}$ and $\eta_{\phi_{\mu},\phi_{\nu}}^{(b,\delta m)}$ are defined by
\begin{eqnarray} 
\eta_{\phi_{\mu},\phi_{\nu}}^{(f,\delta m)} &=& \sum_m \langle \delta m + m | D(i\rho_c^{\frac{1}{2}}) | m\rangle \langle \phi_{\mu} | \delta m + m\rangle 
\langle m| \phi_\nu\rangle\nonumber\\
\eta_{\phi_{\mu},\phi_{\nu}}^{(b,\delta m)} &=& \sum_m \langle \delta m + m | D(-i\rho_c^{\frac{1}{2}}) | m\rangle \langle \phi_{\mu} | \delta m + m\rangle 
\langle m| \phi_\nu\rangle\nonumber\\
&=& (\eta_{\phi_\nu,\phi_{\mu}}^{(f,-\delta m)})^\ast,
\label{eta_2_27_26}
\end{eqnarray} 
while $\varepsilon_l^{(f,\delta m,i)}$ and $\varepsilon_l^{(b,\delta m,i)}$ for $i=1,2,3$ are defined by
\begin{eqnarray}
\varepsilon_l^{(f,\delta m,1)}  &=& 
E_{\phi_{\mu}}  - E_{\phi_\nu} + \varepsilon_k - eV + E_I(1+2q)
+\frac{\hbar \omega_p \delta m}{2},
\nonumber\\
\varepsilon_l^{(f,\delta m,2)}  &=& 
E_{\phi_{\nu}} - E_{\phi_{\mu}} + \varepsilon_k - eV + E_I(1+2q) + \frac{\hbar\omega_p\delta m}{2},
\nonumber\\
\varepsilon_l^{(f,\delta m,3)}  &=& 
E_{\phi_{\nu}} - E_{\phi_{\mu'}} + \varepsilon_k - eV + E_I(1+2q) + \frac{\hbar\omega_p\delta m}{2},
\nonumber\\
\varepsilon_l^{(b,\delta m,1)}  &=& 
E_{\phi_{\nu}} - E_{\phi_{\mu}} + \varepsilon_k - eV - E_I(1-2q) - \frac{\hbar\omega_p\delta m}{2},
\nonumber\\
\varepsilon_l^{(b,\delta m,2)}  &=&
E_{\phi_{\mu}} - E_{\phi_{\nu}} + \varepsilon_k - eV - E_I(1-2q) - \frac{\hbar\omega_p\delta m}{2},
\nonumber\\
\varepsilon_l^{(b,\delta m,3)}  &=& 
E_{\phi_{\mu'}} - E_{\phi_{\nu}} + \varepsilon_k - eV - E_I(1-2q) - \frac{\hbar\omega_p\delta m}{2}.\nonumber\\
\label{varepsilons_6_10_26}
\end{eqnarray} 
Here, $E_I=e^2/(2C_I)$.
In Eq.~(\ref{eta_2_27_26}), $\rho_c$ is the interaction parameter defined by $\rho_c=\alpha_c^2 \sqrt{E_C/ (8E_J)} $ with $E_C=e^2/(2C_r)$.
The translation operator $D(X)$ is defined as $D(X)=\exp[X a^\dagger - X^\ast a]$.
The superscript $f$ of $\eta_{\phi_{\mu},\phi_{\nu}}$ and $\varepsilon_l$ denotes the forward tunneling where the number of \textcolor{black}{quasiparticle}s increases in the superconducting island, while $b$ denotes opposite (backward) tunneling.

Because of Eq.~(\ref{EOM_2_26_26}), $\Gamma^{(1)}(\phi_{i},\phi_{i},\phi_{j},\phi_{j},V)$ can be regarded as the rate of the transition from $|\phi_{j}\rangle$ to $|\phi_{i}\rangle$ caused by the QCR, and is quantitatively studied in the following section. The other $\Gamma$s are also important to describe the dynamics of the KPO.

\section{Probability $p_q$}
\label{Probability p}
We calculate the probability $p_q$ that the superconducting island is in the charge state $|q\rangle$ using the approach developed in Refs.~\cite{Silveri2017,Masuda2025}.
This method assumes that the KPO remains predominantly within the qubit subspace and that elastic quasiparticle tunneling occurs on a much faster timescale than inelastic tunneling processes.
Under these conditions, the charge distribution of the superconducting island can be determined from elastic tunneling alone.

We consider quasiparticle tunneling events that occur when the initial charge number of the superconducting island is $q$, accompanied by a KPO transition $\phi_\mu \rightarrow \phi_\nu$.
Both SIS junctions are assumed identical, with bias voltage $V$ applied to each.
The tunneling rates are denoted as $\Gamma^{+}_{L\leftarrow C}(\phi_{\mu},\phi_\nu,q,V)$, $\Gamma^{+}_{C\rightarrow R}(\phi_{\mu},\phi_\nu,q,V)$, $\Gamma^{-}_{L\rightarrow C}(\phi_{\mu},\phi_\nu,q,V)$, and $\Gamma^{-}_{C\leftarrow R}(\phi_{\mu},\phi_\nu,q,V)$, where the superscript $+$ ($-$) indicates an increase (decrease) in the island charge by one. Here the subscripts $L$, $C$, and $R$ indicate the left superconducting electrode, center superconducting island, and right superconducting electrode, respectively. 
For example, $\Gamma^{+}_{L\leftarrow C}(\phi_{\mu},\phi_\nu,q,V)$ is the transition rate of a quasi-particle tunneling from the superconducting island to the left superconducting electrode.

The rates are calculated using the same technique as for $\Gamma^{(i)}$ in Eq.~(\ref{EOM_2_26_26}), and are given by
\begin{eqnarray}
\Gamma^{+}_{L\leftarrow C}(\phi_{\mu},\phi_\nu,q,V) &=&
\frac{2}{e^2{R_T}} \sum_{\delta m}  \int d\varepsilon_k n_s(\varepsilon_k) n_s(\varepsilon_l) \nonumber\\
&& \times [1-f(\varepsilon_k,T_L)] f(\varepsilon_l,T_C)
|\eta_{\phi_{\mu},\phi_{\nu}}^{(f,\delta m)}|^2 \nonumber\\
\Gamma^{+}_{C\rightarrow R}(\phi_{\mu},\phi_\nu,q,V) &=&
\Gamma^{+}_{L\leftarrow C}(\phi_{\mu},\phi_\nu,q,-V), \nonumber\\
\Gamma^{-}_{L\rightarrow C}(\phi_{\mu},\phi_\nu,q,V) &=&
\frac{2}{e^2{R_T}} \sum_{\delta m}  \int d\varepsilon_k n_s(\varepsilon_k) n_s(\varepsilon_l) \nonumber\\
&& \times 
f(\varepsilon_k,T_R)[1-f(\varepsilon_l,T_C)]
|\eta_{\phi_{\mu},\phi_{\nu}}^{(b,\delta m)}|^2, \nonumber\\
\Gamma^{-}_{C\leftarrow R}(\phi_{\mu},\phi_\nu,q,V) &=&
\Gamma^{-}_{L\rightarrow C}(\phi_{\mu},\phi_\nu,q,-V),
\end{eqnarray}
where $\varepsilon_l=\varepsilon_l^{f,\delta m,1}(V)$ as defined in Eq.~(\ref{varepsilons_6_10_26}).

To clarify which tunneling processes dominantly determine the charge distribution $p_q$, we define:
\begin{eqnarray}
\Gamma_q^{\phi_{\mu}\rightarrow\phi_\nu}(V)&=&\Gamma^{+}_{L\leftarrow C}(\phi_{\mu},\phi_\nu,q,V)+\Gamma^{+}_{C\rightarrow R}(\phi_{\mu},\phi_\nu,q,V)\nonumber\\
&&-\Gamma^{-}_{L\rightarrow C}(\phi_{\mu},\phi_\nu,q,V)-\Gamma^{-}_{C\rightarrow R}(\phi_{\mu},\phi_\nu,q,V).\nonumber \\
\end{eqnarray}
$\Gamma_q^{\phi_{\mu}\rightarrow\phi_\nu}(V)$ represents the net rate of charge change in the superconducting island due to quasiparticle tunneling associated with the KPO transition $\phi_{\mu}\rightarrow\phi_\nu$.

\begin{figure}
\begin{center}
\includegraphics[width=7.5cm]{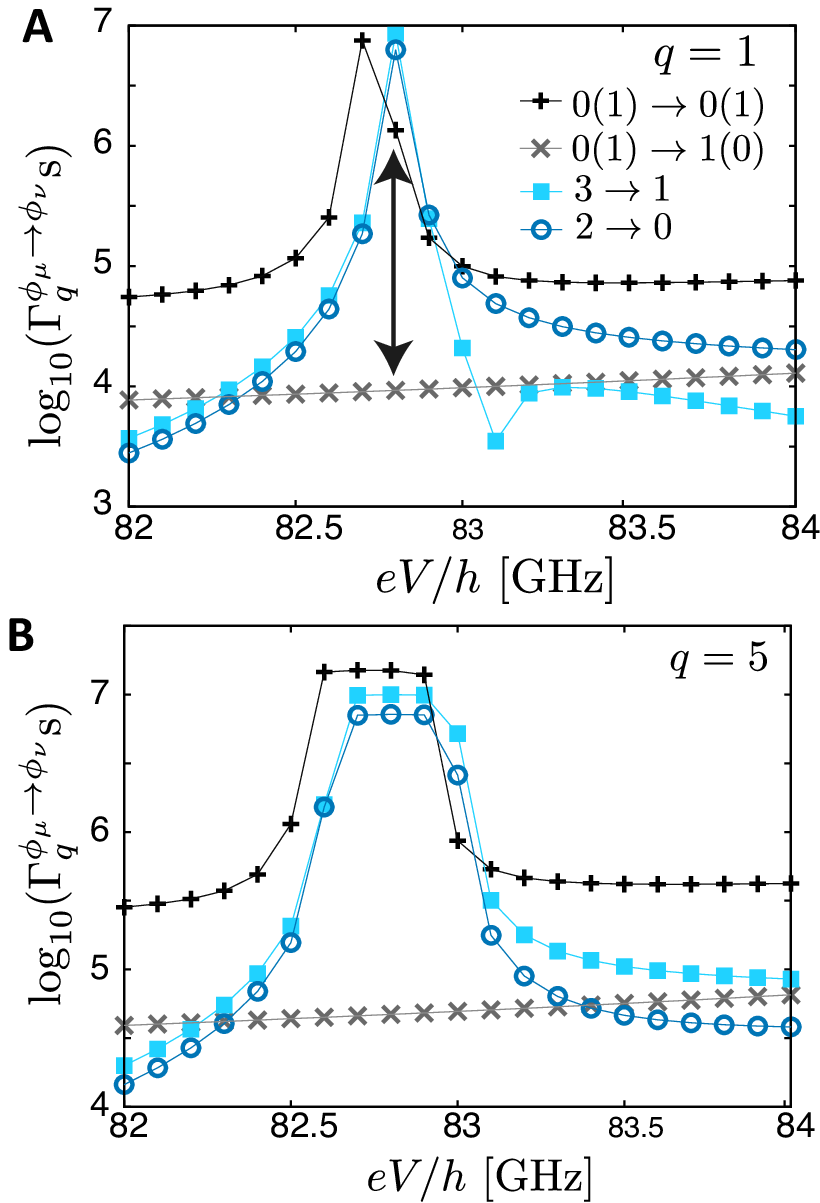}
\end{center}
\caption{
Bias-voltage dependence of transition rate $\Gamma_q^{\phi_{\mu}\rightarrow\phi_\nu}$ for elastic and inelastic KPO transitions for $q=1$ (A) and $q=5$ (B), around the bias voltage regime where two-photon absorption occurs.
The coupling parameter is set to $\rho_c=0.01$.
All other parameters are the same as those used in Fig.~\ref{Gamma1_V_4_2_26}.
}
\label{Gamma_q_plus_V_com_6_10_26}
\end{figure}

To justify the assumption that $p_q$ is determined by elastic tunneling alone, we compare elastic and inelastic tunneling rates.
As will be shown below, the approximation remains valid at the operating point where two-photon absorption provides the dominant cooling mechanism.
Figure~\ref{Gamma_q_plus_V_com_6_10_26} shows $\Gamma_q^{\phi_{\mu}\rightarrow\phi_\nu}(V)$ for elastic and inelastic KPO transitions at $\rho_c=0.01$, around the bias voltage regime where two-photon absorption occurs.  
It is evident that, in certain bias voltage regimes, the rate of elastic tunneling ($\phi_{0(1)}\rightarrow\phi_{0(1)}$) is much faster than inelastic tunneling ($\phi_{0(1)}\rightarrow\phi_{1(0)}$) and heating transitions, which are not shown in the figure.  
We therefore assume that elastic quasiparticle tunneling determines $p_q$.  
Although inelastic tunneling associated with cooling ($\phi_{3(4)}\rightarrow \phi_{1(2)}$) can be comparable to elastic tunneling, it is negligible in this study since the qubit is mostly in the qubit subspace.

The probability $p_q$ is expressed in terms of the rates for tunneling that increase and decrease $q$ as
\begin{eqnarray}
p_q =  \frac{1}{Z} \prod_{q'=0}^{q-1} \frac{\Gamma_{q',\phi_0,\phi_0}^+}{\Gamma_{q'+1,\phi_0,\phi_0}^-},
\label{p_q_6_15_26}
\end{eqnarray}
where $Z$ is a normalization factor, and $\Gamma^\pm_{q,\phi_0,\phi_0}(V)$ are defined by
\begin{eqnarray}
\Gamma^+_{q,\phi_0,\phi_0}(V) &=& \Gamma^{+}_{L\leftarrow C}(\phi_{0},\phi_0,q,V) + 
\Gamma^{+}_{C\rightarrow R}(\phi_{0},\phi_0,q,V),\nonumber\\
\Gamma^-_{q,\phi_0,\phi_0}(V) &=& \Gamma^{-}_{L\rightarrow C}(\phi_{0},\phi_0,q,V)
+ \Gamma^{-}_{C\leftarrow R}(\phi_{0},\phi_0,q,V). \nonumber\\
\label{Gamma_pm_6_15_26}
\end{eqnarray}
In Eq.~(\ref{p_q_6_15_26}), $\Gamma_{q',\phi_1,\phi_1}^+$ and $\Gamma_{q'+1,\phi_1,\phi_1}^-$ could be used instead, as they are approximately equal to $\Gamma_{q',\phi_0,\phi_0}^+$ and $\Gamma_{q'+1,\phi_0,\phi_0}^-$, respectively.

\begin{figure}
\begin{center}
\includegraphics[width=7.5cm]{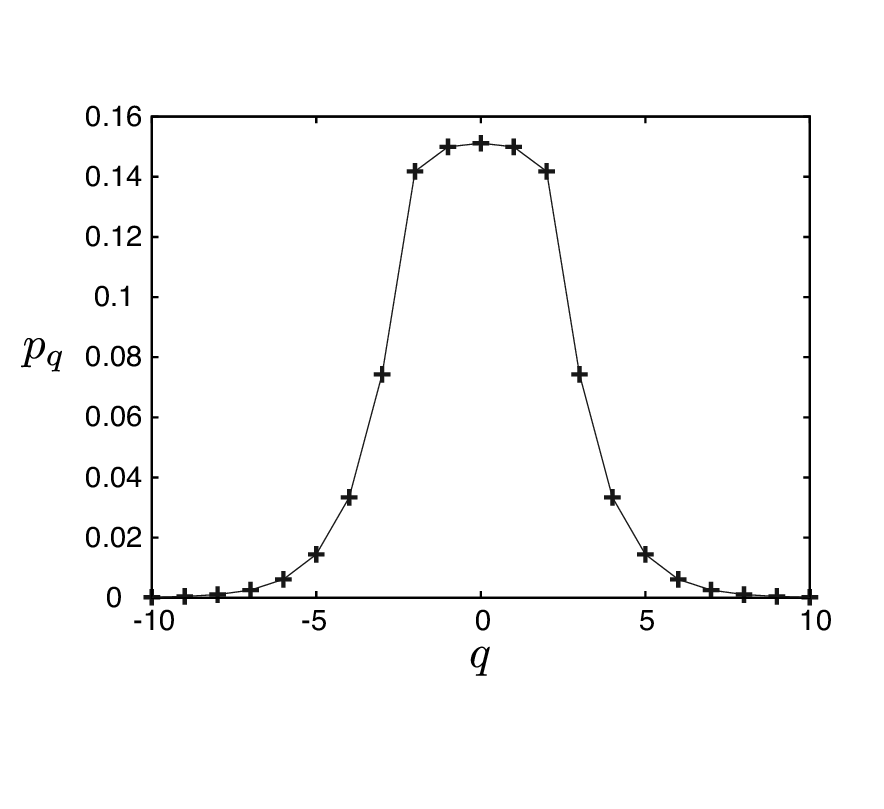}
\end{center}
\caption{
Probability distribution $p_q$ of the superconducting-island charge state $|q\rangle$ as a function of $q$ for $\rho_c=0.01$ for $eV/h=82.8$~GHz.
All other parameters are identical to those used in Fig.~\ref{Gamma1_V_4_2_26}.
}
\label{p_q_8_5_26}
\end{figure}

Figure~\ref{p_q_8_5_26} presents the probability distribution $p_q$ as a function of $q$ at a bias voltage where the two-photon absorption process dominates over other inelastic processes.
In contrast to a SINIS-based QCR, which exhibits an approximately Gaussian distribution of $p_q$~\cite{Silveri2017}, our system displays a noticeable deviation from a Gaussian profile, with a relatively flat region around the peak.
This behavior originates from the sharp structure of the superconducting density of states, which modifies the transition rates between charge states.

\section{Details of time-evolution calculations}
\label{Details of time-evolution calculations}
To simulate the dynamics shown in Fig.~\ref{fig_time_deps_4_13_26}, we include pure dephasing and internal loss in addition to the QCR-induced dynamics derived in Sec.~\ref{Effective QCR-induced dynamics}.
In the absence of the QCR, the density matrix obeys
\begin{eqnarray}
\frac{d\rho^{\rm KPO}(t)}{dt} &=& - \frac{i}{\hbar} [H_{\rm KPO}^{\rm (RF)},\rho^{\rm KPO}(t)]  
+ \frac{\kappa}{2} \mathcal{D}[a] \rho^{\rm KPO}(t)\nonumber\\
&&+ \gamma_p \mathcal{D}[a^\dagger a] \rho^{\rm KPO}(t),
\label{ME_wo_QCR_4_8_24}
\end{eqnarray} 
where $\mathcal{D}[\hat{O}]\rho = 2\hat{O}\rho \hat{O}^\dagger -  \hat{O}^\dagger \hat{O} \rho - \rho \hat{O}^\dagger \hat{O}$~\cite{Puri2020}. Here, $\kappa$ and $\gamma_p$ are the single-photon-loss rate and the pure-dephasing rate, respectively.
The full equation of motion, including the QCR-induced dynamics, is numerically integrated using a fourth-order Runge--Kutta method.

\section{Dependence on tunnel resistance}
\label{Dependence on tunnel resistance}
Because the QCR-induced transition rates scale inversely with the tunnel resistance $R_T$ [Eq.~(\ref{Gamma_2_27_27})], $R_T$ serves as an important tuning parameter that controls both the desired cooling transitions and unwanted phase-flip processes.
Consequently, an appropriate choice of $R_T$ is required to achieve a favorable balance between leakage suppression and preservation of the biased error structure.
To illustrate the role of $R_T$, we examine the dependence of the KPO dynamics on the tunnel resistance.

\begin{figure}
\begin{center}
\includegraphics[width=7.5cm]{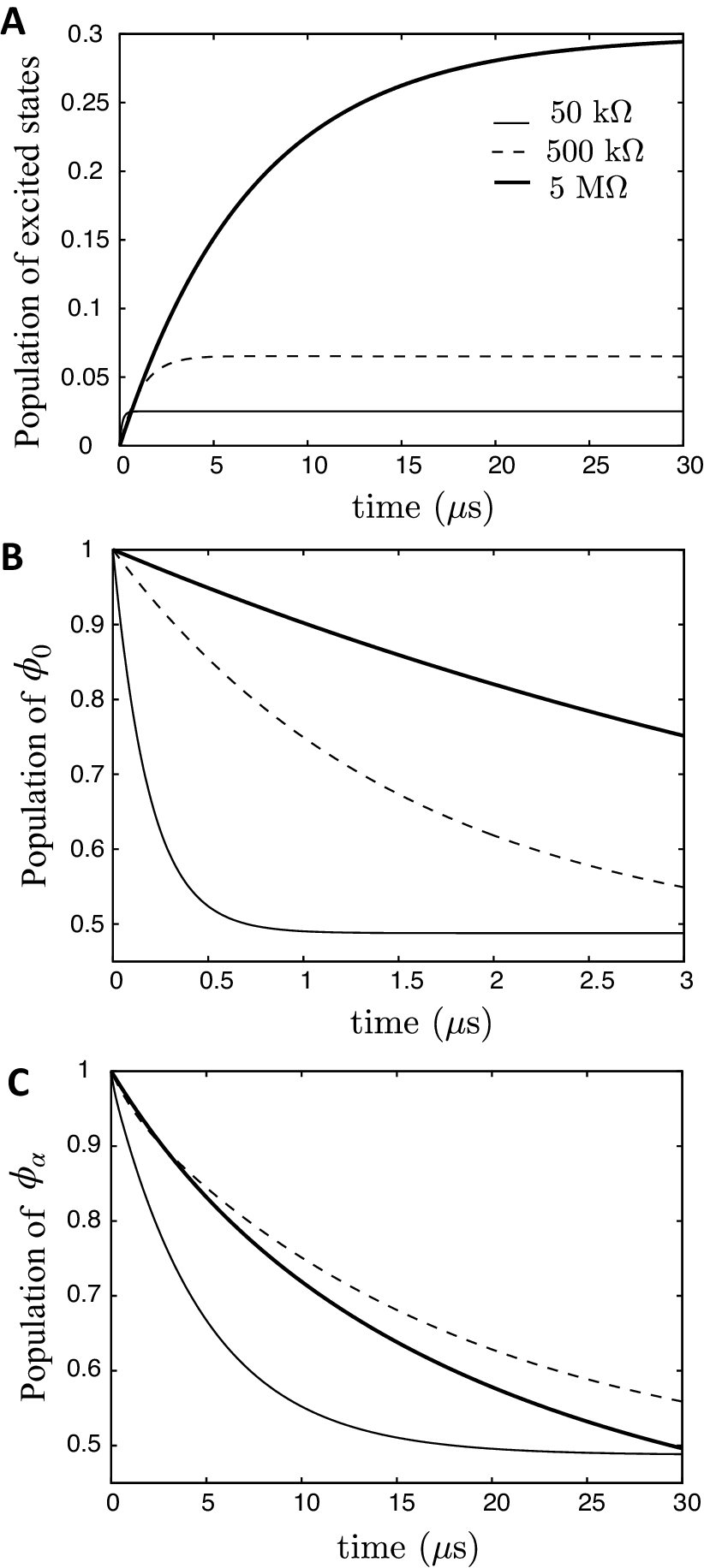}
\end{center}
\caption{
Time evolution of state populations under pure dephasing, internal loss, and a SISIS-based QCR for different tunnel resistances $R_T=50$~k$\Omega$, $500$~k$\Omega$, and $5$~M$\Omega$.
The $R_T=500$~k$\Omega$ data correspond to those shown in Fig.~\ref{fig_time_deps_4_13_26} and are included for comparison.
(A) Population of excited (leakage) states.
(B) Population of $|\phi_0\rangle$.
(C) Population of $|\phi_\alpha\rangle$.
The initial state is $|\phi_0\rangle$ in (A) and (B), and $|\phi_\alpha\rangle$ in (C).
All other parameters are the same as in Fig.~\ref{fig_time_deps_4_13_26}.
}
\label{fig_time_RTdeps_7_9_26}
\end{figure}

We numerically simulate the dynamics of the KPO under the combined effects of pure dephasing, internal loss, and the SISIS-based QCR.
Figure~\ref{fig_time_RTdeps_7_9_26} shows the time evolution of the populations of relevant states.
The initial state is set to $|\phi_0\rangle$ in panels (A) and (B), and to $|\phi_\alpha\rangle$ in panel (C).
We compare the results for $R_T=50$~k$\Omega$, $500$~k$\Omega$, and $5$~M$\Omega$.

As shown in Fig.~\ref{fig_time_RTdeps_7_9_26}(A), decreasing $R_T$ strengthens QCR-induced transitions and thereby enhances the suppression of leakage.
Conversely, for $R_T=5$~M$\Omega$, the QCR is too weak to efficiently remove dephasing-induced excitations, resulting in a substantially larger leakage population.
However, reducing $R_T$ also enhances QCR-induced phase-flip processes, leading to a reduced population of $|\phi_0\rangle$ [Fig.~\ref{fig_time_RTdeps_7_9_26}(B)].
This demonstrates that excessively strong cooling can degrade the preservation of the computational subspace even in the SISIS configuration.

Figure~\ref{fig_time_RTdeps_7_9_26}(C) shows that the population of $|\phi_{\pm\alpha}\rangle$ is considerably less sensitive to $R_T$.
This behavior reflects the biased-noise nature of Kerr-cat qubits, for which bit-flip errors remain much weaker than phase-flip errors.
Nevertheless, excessively small $R_T$ also leads to a noticeable increase in the bit-flip rate.

Taken together, these results indicate that neither excessively large nor excessively small values of $R_T$ are desirable:
large $R_T$ yields insufficient cooling, whereas small $R_T$ enhances unwanted QCR-induced errors.
An intermediate range of $R_T$ therefore provides a favorable compromise between leakage suppression and preservation of the biased error structure.

\end{document}